1 **Formation and disruption of current filaments in a flow-driven**

2 **turbulent magnetosphere**

4 W. W. Liu[1,4], L. F. Morales[1], V. M. Uritsky[2], and P. Charbonneau[3]

6 **Abstract**. Recent observations have established that the magnetosphere is a system of

7 natural complexity. The co-existence of multi-scale structures such as auroral arcs,

8 turbulent convective flows, and scale-free distributions of energy perturbations has

9 lacked a unified explanation, although there is strong reason to believe that they all stem

10 from a common base of physics. In this paper we show that a slow but turbulent

11 convection leads to the formation of multi-scale current filaments reminiscent of auroral

12 arcs. The process involves an interplay between random shuffling of field lines and

13 dissipation of magnetic energy on sub-MHD scales. As the filament system reaches a

14 critical level of complexity, local current disruption can trigger avalanches of energy

15 release of varying sizes, leading to scale-free distributions over energy perturbation,

16 power, and event duration. A long-term memory effect is observed whereby the filament

17 system replicates itself after each avalanche. The results support the view that that the

18 classical and inverse cascades operate simultaneously in the magnetosphere. In the

[1] Space Science and Technology Branch, Canadian Space Agency

[2] Department of Physics and Astronomy, University of Calgary

[3] Department of Physics, University of Montreal

[4] Also at College of Electronic Information, Wuhan University

19     former, the high Reynolds-number plasma flow disintegrate into turbulence through

20     successive breakdowns; in the latter, the interactions of small-scale flow eddies with the

21     magnetic field can self-organize into elongated current filaments and large-scale energy

22     avalanches mimicking the substorm.

## 24     1. INTRODUCTION

25     Energy release in the magnetosphere manifests itself as geomagnetic and auroral

26     perturbations. Detailed analyses have shown that these perturbations follow the so-called

27     scale-free distributions (*Consolini*, 1997; *Lui et al.,* 2000; *Uritsky et al*, 2002; 2009;

28     *Kozelov et al.,* 2004). For instance, *Uritsky et al.* (2002) found that the probability density

29     function over auroral brightness integrated over space and time (called *E*) has a power-

30     law form $E^{-\alpha}$, where $\alpha$ is a constant. What scale-free distributions mean in the context

31     of magnetospheric physics has drawn considerable interest of late. One interpretation is

32     that the active magnetosphere is in a state of self-organized criticality (SOC); energy

33     releases in a SOC state can have different sizes, but the governing physics is the same. A

34     number of theoretical and simulation studies have been carried out, in which scale-free

35     distributions of magnetospheric perturbations were reproduced (*Chapman et al.,*1998;

36     *Klimas et al.,* 2000, 2004; *Uritsky et al.,* 2001; *Valvidia et al.,* 2003; *Liu et al.,* 2006;

37     *Valliere-Nollet et al.*, 2010).

38     While scale-free dynamics may be mathematically elegant and conceptually appealing,

39     a deeper inspection brings us to an apparent contradiction: The structures that are

40     associated with or responsible for energy release do not follow scale-free statistics. It is

41    well-known that active aurora is dominated by discrete arcs, and the disruption of

42    equatorward arcs lies at the heart of auroral substorm onsets (*Akasofu*, 1964). The

43    relationship of the disruption to propagation of substorm perturbations in the

44    magnetosphere was recently elaborated by *Donovan et al.* (2008). *Knudsen et al.* (2001)

45    performed a quantitative study of the thickness of the 557.1 nm green line excited by 1-

46    10 keV electrons and found a centered distribution with a mean thickness of ~18 km.

47    Embedded in the Knudsen distribution are finer-scale arc populations with thicknesses ~1

48    km (*Partamies et al.,* 2010), ~100 m (*Trondsen et al.,* 1998) and ~10 m (*Maggs and*

49    *Davis,* 1968). Although the structuring of auroral arcs has not been completely resolved

50    as an observational problem, it is generally agreed that the scale distribution of aurora is

51    not a smooth continuum but has multiple peaks. How do we reconcile the discrete

52    structuring of arcs with scale-free dynamics of energy release? The incongruity of this

53    question led *Knudsen et al.* (2001) to assert that "the arc width spectrum argues against

54    the notion of a turbulent cascade of energy from larger to small scales."

55        The formation of auroral arcs is by no means a settled question. As will be elaborated

56    in a separate study, arcs in the Knudsen population typically have longitudinal lengths of

57    several thousand km, which maps to a scale comparable to the size of the magnetosphere.

58    Moreover, the lifetime of these arcs is typically well over 1 min, which is approximately

59    the Alfven transit time. These properties hint strongly that these arcs are regulated by the

60    magnetosphere. While processes in the auroral acceleration region 1-2 Re above Earth

61    can explain the observed thickness of Knudsen arcs (e.g., *Borovsky* (1993)), it is unlikely

62    that long arcs are formed without any organization on the part of the magnetosphere, for

63    otherwise one would be forced to concoct theories why an aurora arc align itself so

64    perfectly over the magnetospheric scale without the magnetosphere playing a role. From

65    the temporal point of view, auroral features lasting longer than the Alfven transit time

66    must maintain some equilibrium with equivalent features in the magnetosphere. Last but

67    not least is the 18-km average thickness. At the approximate 67° magnetic latitude where

68    the Knudsen population was sampled by the CANOPUS all-sky camera in Gillam, the

69    latitudinal mapping factor has the order ~50; a 18-km thick arc should map to the central

70    plasma sheet (CPS) as a filament ~900 km in width. In comparison, a 10 keV proton in a

71    20-nT magnetic field has a gyroradius ~500 km. Therefore, while the cross-tail length of

72    an arc mapped to the magnetosphere is definitely of the MHD scale, its width is likely

73    controlled, in part, by dissipation effects on the ion scale.

74    Hence, if we accept the premise of magnetospheric origin for auroral arcs, as

75    observations compel us to, we must deal with conceptual problems on several fronts. One

76    has to do with the metastability of arcs. By metastable we mean that the arcs maintain a

77    steady form for a period longer than the Alfven transit time (~1 min for the CPS). Under

78    this condition, one would be tempted to view arcs as a characteristic solution of the

79    quasistatic convection problem. However, even in the latest edition of the Rice

80    Convection Model (e.g., *Lemon et al.,* 2004), arc-like solutions do not exist; neither do

81    these structures arise naturally in global MHD simulations. In fact, the actual condition of

82    the magnetosphere poses an even more confounding problem. In-situ observations of

83    plasma flows in the plasma sheet paint a system that is rather turbulent, with the rms

84    speed much larger than the average speed (*Angelopoulos et al.*, 1992; 1999; *Borovsky et*

85    *al.*, 1997; *Borovsky and Funsten,* 2003). How can metastable, arc-like structures survive

86    in, let alone be produced by, a turbulent magnetosphere? Little consideration has been

87    given to this question in the literature. The stationary Alfven wave theory of *Knudsen*

88    (1996) predicts arcs with thickness a few times the electron inertial length in the topside

89    ionosphere (~ 1 km), but requires some ionospheric irregularity (i.e., proto-arc) to anchor

90    the resulting structure. Field-line resonances (FLRs) (*Southwood,* 1974; *Chen and*

91    *Hasegawa,* 1974) give arc-like structures, and observations showed that some arcs indeed

92    oscillate at ULF frequencies predicted by FLR theories (e.g., *Xu et al.*, 1993; *Liu et al.,*

93    1995). However, for those arcs which oscillate, the fluctuation is typically a small

94    fraction of the overall brightness (e.g., *Uritsky et al.*, 2009). We are still left with the task

95    of explaining the dominant non-oscillating part of the arcs.

96      The brief review above points to significant gaps in our knowledge of the relationship

97    between magnetospheric structures and dynamics of energy release usually associated

98    with the collapse of these structures. Of particular interest are the following questions:

99    How do metastable arc-like structures form in a turbulent magnetosphere? What makes

100   these structures collapse? What is the distribution of energy release from the collapse? At

101   present we lack a clear program to formulate answers to these questions, a task we

102   embark upon from the point of view of nonlinear multi-scale coupling.

103     As a first step, we develop a new framework whose salient properties are investigated

104   with a simplified model. As a point of departure, we begin with a magnetosphere in a

105   state of weak turbulence (in the sense that the flow speed is much smaller than the speeds

106   of MHD modes).  We track the change of the magnetic field frozen in the flow and

107    observe the current structures resulting from the random shuffling of field lines. In a

108    surprising twist, we will show that the resulting current distribution does not have the

109    uncorrelated random appearance of its turbulent driver but exhibits elongated filamentary

110    structures reminiscent of arcs. In section 2, we give the basic outline of the theory, as

111    well as key assumptions of the model. In section 3, we present simulation results from

112    select runs of the model, including time series of energy avalanche, probability density

113    functions of energy release, and morphology of representative current distributions. In

114    section 4, we discuss the implications of the results in the context of multiscale

115    magnetospheric dynamics and propose an interpretation of magnetospheric dynamics

116    based on the idea of natural complexity.

117

118    **2. THEORY**

119    Bright auroral arcs are generated by energetic electron precipitation and associated

120    principally with upward field-aligned currents (FACs) denoted as $j_{\parallel}$. By virtue of current

121    continuity, a FAC is related to the magnetospheric current $\mathbf{j}_{\perp}$ perpendicular to magnetic

122    field as

123    $$j_{\parallel} = -B_i \int \frac{\nabla \cdot \mathbf{j}_{\perp} \, ds}{B} \qquad (1)$$

124    where $ds$ denotes integration along a field line, and the subscript $i$ denotes value at the

125    ionospheric foot print. For metastable arcs with lifetime longer than the Alfven transit

126    time, (1) implies that, after adjustment for mapping, auroral structures associated with $j_{\parallel}$

127 should correspond to similar structures in $\mathbf{j}_\perp$. *Elphinstone et al.* (1991) showed that there

128 is indeed a close correlation between aurora arcs observed by the Viking UV imager and

129 cross-tail current in the magnetosphere. In this paper we direct our attention to how arc-

130 like structures can be formed as the magnetospheric **B** field evolves in a turbulent

131 convection. It bears further notice that the smaller the scale length of $\mathbf{j}_\perp$, the larger the

132 magnitude of $j_\parallel$, explaining why thin arcs tend to be brighter.

133  Figure 1a is a representation of the magnetosphere. The plasma sheet situated on the

134 night side is generally considered as the source of discrete aurora arcs in the oval.

135 Particularly, the equatorward arcs sampled by *Knudsen et al.* (2001) map mostly to the

136 central plasma sheet (CPS) located earthward of 15 Re. In Figure 1b, the CPS is

137 abstracted as a collection of discrete flux tubes identified by their foot points through

138 equatorial plane. In a weakly turbulent magnetosphere, the foot prints undergo slow

139 quasi-random motions (by quasi-random we mean that the motions appear random and

140 uncorrelated beyond the correlation length of the turbulent field). To simplify the

141 problem and make the salient points more transparent, we take the field lines as straight.

142 This approximation removes field line curvature, which accounts for a large part of the

143 perpendicular current that feeds the FAC in (1), hence limiting the literal use of the model

144 in its present form. This caveat notwithstanding, we expect that the salient features

145 emphasized by the present study, namely, the relationship between current filaments and

146 turbulence, as well as the scale-free nature of energy release, should survive this

147 approximation. At this point, the objective of our treatment is to substantiate the 148    plausibility of an idea rather than simulating the behavior of an actual system.

149    We use the magnetic field $B_z$ as the primary variable. At the start of simulation, $B_z$ is

150    initialized as a linearly decreasing function of x. The electric field in the plane is given by

151    $$\mathbf{E} = -\mathbf{v} \times \mathbf{B} + \eta \nabla \times \mathbf{B} \qquad (2)$$

152    where η is the plasma resistivity. *Lui et al.* (2007) analyzed the Vlasov-averaged version

153    of generalized Ohm's law in a neutral sheet crossing event observed by the Cluster

154    satellites and found that the resistivity term accounted for most of the deviation from the

155    ideal MHD condition, with a magnitude comparable to the E and v×B terms individually.

156    For the typical parameters given in the event of *Lui et al.* (2007) and assuming a current

157    sheet thickness 1000 km, we find that η has an order of magnitude ~$10^{11}$ m$^2$/s, which is a

158    significant value. Formally the resistivity term written by *Lui et al.* (2007) represents the

159    effects of electromagnetic turbulence and was found to be predominantly dissipative (i.e.,

160    $\mathbf{j} \cdot \mathbf{E} > 0$). This finding is consistent with the following interpretation: As the shuffling of

161    field lines create more and more complex structures in $B_z$, electromagnetic turbulence on

162    the ion scale and below is excited. These turbulent excitations are a conduit which

163    transfers energy from the magnetic field to thermal energy of particles. In this manner,

164    the dissipation prevents the formation of excessively sharp structures.

165    Faraday's law, coupled with the incompressibility condition, gives the rate of change

166    of the magnetic field as

167    $$\frac{\partial B_z}{\partial t} = -\mathbf{v} \cdot \nabla B_z + \eta \nabla^2 B_z \qquad (3)$$

168    Equation (3) is solved on a two-dimensional coupled lattice. Simulations are performed

169    on a 256×256 grid. If the size of the physical system, is 20 $R_E$ ×20 $R_E$ , one grid spacing

170    $\Delta$ at the 256×256 resolution has the approximate length 500 km, comparable to the ion

171    gyroradius cited earlier. Physics below this scale is represented by kinetic dissipation

172    through $\eta$ .

173    We take **v** as given. At each time step, the velocity is prescribed randomly at each

174    node. In a realistic turbulence, flow velocities become independent only beyond a finite

175    correlation length. The above implementation, adopted mainly for its convenience,

176    implies that the correlation length is less than the grid spacing. In truth, this condition

177    does not typically apply to Earth's magnetosphere. *Borovsky and Funsten* (2003), for

178    example, estimated that the correlation length of magnetospheric turbulence is of the

179    order 1-2 $R_E$ . As these authors pointed out, the size of the CPS (whose thickness is also

180    a few $R_E$ ) is comparable to the inferred correlation distance, giving a sort of "turbulence-

181    in-a-box" which deviates from the classical turbulence with well-separated injection,

182    inertial and dissipation scales. To bring clarity to the problem at hand, we defer this detail

183    for future consideration and assume that the turbulence following a power-law

184    distribution of energy density, $\varepsilon(k) \propto k^{-a}$ , where $\varepsilon(k)$ is energy per wave number $k$.

185    (The classical Kolmogorov turbulence has $a = 5/3$ .) The velocity at scale $k$ is $v_k \propto k^{\frac{1-a}{2}}$ .

186    It can be shown that the first term on the right-hand side of (3), which drives the

187    formation of structure in $B_z$ , varies as $k^{\frac{3-a}{2}}$ , whereas the dissipation term varies as $k^2$ . If

188    the driving turbulence has $a < 3$ , equation (2) predicts that small-scale structures grow

189  faster than large-scale ones. Since the current density at scale $k$ is $j_k \propto kB_k \propto k^{\frac{5-a}{2}}$, the

190  process will quickly lead to the formation of small-scale current structures. Eventually,

191  the dissipation $\eta$ kicks in and the formation of structures stops at a scale $k_c \propto \eta^{-\frac{1+a}{2}}$.

192  Because of the faster growth of small-scale structures, it is a reasonable first

193  approximation to retain only the uncorrelated flow components at the scale $\Delta$ and below;

194  this flow component is a fraction of the observed flow speed at any given point.

195  Effectively, our present implementation implies that flow components at scales larger

196  than $\Delta$ do not contribute significantly to the formation of current structures. By the same

197  token, the velocity fields between successive time steps are also uncorrelated and

198  prescribed randomly.

199  As the magnetic field evolves in accordance with (3), more and more complex

200  structures form, and the current density increases.  When the local current density

201  exceeds the starting current by a factor $M$, we assume that some form of current-driven

202  instability takes place, and the current distribution is relaxed with a certain amount of

203  energy released. Observationally, the cross-tail current has been observed at values as

204  high as 100 $\mu A/m^2$ (*Asano et al*., 2003; *Nakamura et al.,* 2010), while the quiet-time

205  current density in equatorial plane has the order of 1 $\mu A/m^2$. In our simulation, we have

206  used $M = 2 - 20$ as the instability threshold. Once an instability occurs, we assume that

207  it reduces the local current density to zero. This means that, after the instability, the

208  unstable node and its four nearest neighbors (labeled 0-4) have the same magnetic field

209  equal to the 5-point average before onset, viz, $\langle B \rangle = (B_0 + B_1 + B_2 + B_3 + B_4)/5$.  This

210   procedure conserves magnetic flux and releases an amount of energy equal to

211   $$\Delta E = \frac{1}{2\mu_0} \sum \left( B_i - \langle B \rangle \right)^2 \qquad (4)$$

212   where the sum is over all nodes on the grid.

213   As in *Liu et al.* (2006), a fraction $\delta$ of the energy release goes into Alfvén waves to

214   excite aurora. The rest, $(1-\delta)\Delta E$, stays in the magnetosphere. We make the simple

215   assumption that the retained energy release feeds a plasma flow that blasts out radially

216   from the unstable node. The velocity on the four nearest neighbors has the magnitude

217   $v_b = \sqrt{(1-\delta)\Delta E / 2\rho}$, where $\rho$ is the plasma mass density. The effect of the blasts on the

218   magnetic field is solved through (3). Once the system is settled, we implement the next

219   iteration of the turbulent **v**. A free boundary condition is imposed in the simulation runs;

220   that is, when an avalanche hits the boundary, the energy freely exits the system without

221   any impediment.

222   *Takalo et al.* (1999) studied a coupled-lattice model which at first glance looks similar

223   to ours. A close examination indicates that the two models invoke different physical

224   assumptions. We note the following distinctions in our model: 1) The full induction

225   equation is solved, rather than assuming a source function generating magnetic flux. This

226   allows a direct link to magnetospheric turbulence. 2) The magnetic resistivity is a

227   constant, rather than a function of local current and plays a different role in our model. It

228   can be shown that, if there is only resistivity and no flow, the solution of (2) is simply the

229   decay of the initial $B_z$, without any emergent complexity. It is the turbulent **v** (which,

230   through its product with **B**, constitutes the nonlinearity in our model) that leads to the

231    formation of structures and release of energy; the role of $\eta$ is merely to dissipate energy

232    on the sub-MHD scale. In Takalo et al. (1999), the hysteresis of $\eta$ was the nonlinearity

233    responsible for the resultant complexity. 3) Energy partition in our model is more

234    realistic, with particle heating associated with $\eta$, bulk flows associated with v, and energy

235    flux to the auroral ionosphere associated with the partition of (3). In *Takalo et al.* (1999),

236    only particle heating was present.

237

## 238    3. RESULTS

239    We have run the model under different combinations of parameters. These runs

240    showed a consistent general pattern in terms of structure formation, avalanche, and

241    statistical distributions. In this section, we present samples of the simulation runs to

242    highlight some of the more interesting aspects of this pattern. The dimensionless

243    parameters for these runs were chosen to be $M = 2.5$, $\eta = 10^{-3}$, $v_{\mathrm{rms}} = 10^{-6}$, and

244    $\delta = 0.1$. The choice of parameters was verified *a posteriori* to give filamentary structures

245    with thickness between 1 and 10 $\Delta$, the estimated width of mapped arcs suggested by our

246    previous calculation. More extended analyses and discussion of our model for a broader

247    range of parameters will be reported elsewhere.

248

### 249    3.1. Energy avalanches and self-organized criticality

250    Figure 2 gives the time series of total lattice energy and total liberated energy (namely

251    the sum of (4) over all active nodes) from the coupled lattice over $4 \times 10^6$ iterations of a

252 particular run. For the first $2.5 \times 10^6$ iterations, the system slowly approaches a critical

253 state, as there is an increasing trend of the total magnetic energy stored on the lattice.

254 Afterwards, the system settles on a statistically stationary state, where the average energy,

255 as well as other statistical properties, does not change with time. Whether this state

256 represents a self-organized criticality is a technical matter for future consideration, what

257 is clear is that, once driven into this state, the system spontaneously slips into energy

258 avalanches of varying sizes.

259   Figure 3 shows a typical avalanche in detail. From a lull of no active node, the

260 avalanche starts abruptly, reaching its peak power in a dozen or so iterations. The initial

261 onset of avalanche removes a large amount of free energy from the system, but the

262 system is not completely relaxed, with unstable current structures forming in neighboring

263 nodes that led to further avalanches and secondary peaks of energy release. It takes ~10

264 times longer than the initial peak release for the system to settle, and free energy to be

265 completely removed.  This pattern is similar to the profile of an aurora substorm; that is,

266 the initial expansion phase that is typically the brightest and lasts a few minutes, followed

267 by up to 1 hour of recovery phase where auroral brightness undergoes ebbs and flows

268 before finally dying down.

269   It is noted that, in order to reach a SOC-like state, the system has to be driven slowly

270 (in comparison to the rate of avalanche), and the driver itself is statistically stationary.

271 Neither condition is necessarily fulfilled in the actual magnetosphere. Therefore, Figures

272 3 and 4 represent a theoretical limit that may not be perfectly realized but is instructive in

273 terms of providing insight on how intermittent energy release can result from persistent

274    actions of a turbulent flow.

275

276    **3.2. Probability density distributions**

277    In Figure 4, probability distribution functions of total energy release (*E*), event

278    duration (*T*), and peak power (*P*) are presented. The sample consists of 8676 avalanches.

279    All PDFs are fit to a power law $X^{-\alpha}$, represented by the red line through the

280    corresponding histograms in Figure 5. A visual inspection confirms that distributions of

281    the three parameters have excellent fits to the power laws. Table 1 lists the power law

282    exponents obtained for two different lattice sizes: 128×128 and 256×256. We conclude

283    from the table that the results shown in Figure 5 are statistically robust based on the

284    convergence of α.

285    Due to the approximations made in the current implementation of the model, we do not

286    make direct comparisons of the power-law exponents obtained through simulation to

287    those estimated from real data. It is, however, interesting to note that the power exponent

288    $\alpha_E = 1.14$, for example, is identical to that obtained by *Liu et al.* (2006) obtained through

289    a different approximation of the CPS dynamics.

290

291            Table 1. Simulations parameters and results for the PDF's of avalanches.

| N | $\alpha_E$ | $\alpha_P$ | $\alpha_T$ |
|---|---|---|---|
| 128 | 1.15±0.03 | 0.97±0.06 | 1.41±0.05 |
| 256 | 1.15±0.02 | 1.09±0.06 | 1.37±0.05 |

292

293

### 3.3. Current filaments

294    Figure 5 shows four plots of the current density distribution taken at random points of

295

296    a simulation run. The current density is calculated as $\mathbf{j} = \hat{\mathbf{z}} \times \nabla B_z$. In order to highlight the

297    filamentary current structures, we use a form of contour plot to identify nodes where

298    there is an enhancement of current magnitude, without regard to direction. By connecting

299    the dots, we get a sense of the overall structure of the current distribution. Also, to see the

300    relationship between current distribution and energy release in an avalanche, we plot on

301    the right-hand side of the current distribution the avalanche event in which it found itself,

302    with the arrow indicating the moment when the current distribution was collected.

303    As indicated earlier, the driver to the system is a turbulent flow field that is completely

304    uncorrelated and random on the coupled lattice. It would not be unreasonable to suppose

305    that the current distribution that results should be similarly uncorrelated and random. The

306    actual results defy this expectation. The common feature of the four plots is that the

307    current distribution is highly filamentary, with the length of the filament much greater

308    than the width. In detail the four plots differ, determined largely by their phasing in

309    relation to the energy release at the moment.

310    In general, we expect that a highly structured current distribution should presage a

311    major energy release event, as there is more energy contained in such a configuration.

312    This expectation is largely borne out in Figure 5. Figure 5d has the most complex

313    structuring, with well-defined system-wide filaments. The current distribution is indeed

314    found to be just before the onset of a large secondary peak in an avalanche. Next in level

315    of complexity is Figure 5c. The current distribution in this case is collected between two

316    secondary peaks, as the system was rebuilding free energy for a significant release. The

317    current filaments are weaker than Figure 5c, and there is a new morphological feature

318    which we call patches, marked as hatches in the middle. Further down the scale of

319    complexity comes Figure 5a, where the current distribution is collected from the

320    downward slope of an energy peak. There is a further weakening of the filaments to be

321    barely visible. Figure 4b shows the current distribution collected right at an energy peak.

322    As expected, it is the least structured of the four plots, as the current filaments have

323    practically disappeared. Replacing them are the prominent patches in the middle. We do

324    not have an answer as to why current patches seem more stable than filaments and leave

325    it as a topic for future investigation.

326       It is interesting to note that the four avalanches in Figure 5 were collected at random.

327    One might expect that the current distributions should have no semblance to each other,

328    as each was rebuilt after the system was cleared of free energy, and there should be no

329    long-term memory effect. However, when we inspect the underlying current distributions

330    for the four events, it is clear that they have a significant degree of similarity. Despite

331    waxes and wanes of the current density, and the presence or absence of patches, the

332    overall pattern is slanted at a ~45° angle to the cross-tail line; even the number of

333    filaments does not seem to vary greatly. Hence the system does retain memory. After a

334    more careful observation of the current distribution, we offer the explanation as follows:

335    Once the general pattern of current distribution is formed, randomly at first, in the build-

336  up phase of a simulation run, it cannot be completely erased by an avalanche. Just as in

337  Figure 5b, at the peak of energy release, there are still remnants of the filaments that

338  preceded the event. Then, as the system enters into the next period of energy buildup, the

339  surviving current enhancements serve as the seed to rebuild a current distribution similar

340  to the previous one. The reason is that the current increment per iteration is proportional

341  to the local current density, according to (2). Thus, the surviving current enhancements

342  have the advantage, and the probability of recurrence of the initial distribution is high,

343  even though the driver is random. In a manner of speaking, this behavior is not

344  fundamentally different from the fact that facture tends to happen where the bone has

345  already been broken before or an earthquake is more likely to hit where there is already a

346  fault.

347  To confirm this explanation, we show in Figure 6 the results from a different run of the

348  model. The current distributions just before and after an avalanche are plotted. As our

349  argument above implies, this run initialized a different current pattern from Figure 5.

350  Furthermore, the avalanche did remove energy from the coupled lattice but did not

351  completely erase the underlying pattern, as the current distribution after the avalanche

352  (Figure 6b) is essentially a weakened facsimile of that before the avalanche (Figure 6a).

353  While a first glance at Figure 5 may suggest that the highly structured current

354  distribution is incongruent to the smooth and scale-free energy releases in Figure 4,

355  further reflection indicates that the two can be reconciled. For argument's sake, suppose

356  the system before disruption has $n$ current filaments. Suppose further that the system is

357  near criticality everywhere, and the ensuing avalanche causes all filaments to disrupt, the

358     so-called system-wide discharge. The total energy release under this scenario would have

359     a normalized value $n$. However, it is also possible that only half of the filaments are near

360     criticality, yielding a release of $n/2$. We can follow this logic to the case where only one

361     filament is near criticality, with energy release equal to 1. In fact, it is possible that

362     avalanches occur only in part of a filament, leading to releases that are any fractions of

363     unity. It is also reasonable to suppose that, in a system without built-in preference and

364     selection effect, the smaller the event the higher the probability. For this reason, we

365     expect that the probability density function increases monotonically toward the small

366     releases, although we cannot quite predict that the specific form should be power-law

367     without further analysis or actual simulation.

368

369     **4. DISCUSSION**

370     Filamentary structures are very common in nature. From the cosmic microwave

371     background, to mass distribution in galaxies, to active regions involved in solar flares, to

372     seismic faults, we find matter or energy concentrated in elongated, asymmetric forms.

373     While physics responsible for these phenomena certainly vary, that different physics give

374     rise to similar structures has been cited by many as a sign of universal laws which we do

375     not quite yet grasp but could well exist to govern how complex systems appear and work.

376     Studying aurora and the underlying magnetospheric system from this perspective is an

377     example of this search for potential universality.

378     As an interesting side note, one cannot escape noticing a similarity of auroral

379     phenomena to the seismic system. The distribution of earthquake energy (the Richter

380    Scale) has the scale-free power-law form, whereas the scale distribution of earthquake

381    faults is certainly centered, just like aurora arcs. In the literature, terms such as

382    magnetoseismology and substorm epicenter are seeing regular use. Admittedly, there are

383    areas where aurora and earthquakes differ; for example, seismic faults form mostly along

384    the boundaries of different tectonic plates, whereas aurora arcs can form in a medium that

385    is homogeneous. Nonetheless, the co-existence of centered scale distribution and scale-

386    free energy distribution in both phenomena point to the possibility of a multiscale

387    coupling that features both turbulence and self-organized criticality.

388        The foremost concern of this study was the relationship between magnetospheric

389    turbulence and filamentary current structures which, as we have argued, must underlie

390    metastable auroral arcs. The model we used to establish this potential relationship was

391    simple and should not be used literally to describe the actual magnetospheric physics.

392    However, the salient point concerning the formation of filaments in a totally random

393    flow field is something that transcends the various approximations. What we did in this

394    study was to bring unity to several seemingly unrelated, even contradictory features. We

395    started with a constant (i.e., structureless) current distribution. We drove the system with

396    a completely random flow field. We yielded highly filamentary current distributions from

397    the primordial uniformity. And, finally, we found that the energy release from the

398    filaments is scale-free, returning to a lack of structure many take as a sign of universality.

399    The simplicity of the model with which we unified the disparate strands should be

400    considered a strength, rather than weakness in this regard.

401        Looking forward, there are several aspects of the model that need improvements. We

402  cite a few that are receiving current attention. Magnetic field lines are strongly curved in

403  equatorial plane, so much so that field line curvature **c** can dominate the current density

404  $\mathbf{j} = \mu_0^{-1} \nabla \times \mathbf{B} = \hat{\mathbf{b}} \times \nabla B + B \hat{\mathbf{b}} \times \mathbf{c}$ . In this study, only the first term was considered.

405  Incorporation of the curvature term requires a two-dimensional or field-line integrated

406  model. We anticipate that many of the salient features of the interplay between turbulence

407  and magnetic field should persist in the more realistic implementations, as a turbulent

408  flow would distort the shape of a field line much in the same way as it transports it.

409  We are also looking at a more realistic prescription of **v**. Turbulent flows are to be

410  specified with arbitrary correlation time and length. In this paper we considered only the

411  extreme case of zero correlation time and correlation length. It will be interesting to see

412  how the results might change when the driver maintains a finite correlation in space and

413  time.

414  Ultimately, the turbulent flow **v** should be given self-consistently, rather than specified

415  externally. Just like the kinematic theory of solar dynamo establishes that it is *possible* to

416  generate magnetic field in the convection zone, and it takes a dynamic theory to know

417  exactly how a dynamo works, a central task facing us is to integrate **v** into the model as a

418  co-variable. There are two possible sources of **v**. One is through magnetic reconnection

419  in the tail; the turbulence could be a result of reconnection itself or of the interaction of

420  the flow with local plasma (e.g., *Liu* (2001)). Another possibility is that the flow is the

421  product of local instability. In the latter connection, it is useful to envisage an integration

422  between the present model and the model developed by *Liu et al.* (2006) and *Vallièrès-*

423  *Nollet et al.* (2010) (called LVN). These authors took the pressure (internal energy) as the

424    primary variable, and increased it deterministically to simulate the energization of the

425    plasma sheet in the growth phase. Noting that the current density is related to the pressure

426    gradient by $\mathbf{j} = \mathbf{B} \times \nabla p / B^2$, they made a node topple when $|\nabla p|$ exceeded a prescribed

427    limit. The only random factor in LVN is the energy partition ratio δ; yet scale-free

428    avalanches were a defining characteristic of this system. As mentioned before, the slope

429    of the energy distribution from our model was identical to that predicted by the model of

430    *Liu et al.* (2006). This could mean that scale-free distributions are not sensitive to the

431    choice of primary variable or driver. In its current implementation, the LVN model

432    redistributes all the released energy to neighboring nodes as internal energy (pressure). A

433    modification can be attempted so that the free energy is redistributed into flow $\mathbf{v}$ (as we

434    did with the present model), which can serve as the flow driver to the magnetic field. For

435    an incompressible fluid, the flow would change the pressure distribution through the

436    equation $\partial p / \partial t = -\mathbf{v} \cdot \nabla p$ , which can be solved in much the same way as (3). This

437    approach would maintain the self-consistency between $p$ and $B_z$, as both evolve in time.

438    Despite the various limitations of our model, it is not entirely premature, given the

439    results here and in some of the references, to sketch out a complexity perspective of

440    magnetospheric dynamics, including the nature of substorms. The enunciation of this

441    perspective is not meant to be the final words on the question, as evidence so far has been

442    sketchy, nor a repudiation of other points of view, which all have their basis in facts and

443    logic. Rather, we intend it to be an injection of new ideas that should help broaden our

444    perspective. Key to our outlook are four aspects which merit greater attention: 1)

445 hysteresis, 2) energy storage in multiscale structures, 3) scale-free avalanches associated

446 with the collapse of multi-scale structures, and 4) insensitivity to "triggers." We discuss

447 each in turn, highlighting, where applicable, differences from the traditional view of

448 substorm.

449    Hysteresis (also known as irreversibility) means that in a properly constructed phase

450 space, a system's path of evolution is different from point A to B, as compared to B to A.

451 The area enclosed by the A→B→A loop is usually proportional to a physical quantity

452 (e.g., energy) that is irreversibly released. For store-and-release processes such as the

453 substorm, hysteresis must exist so that the system can accumulate energy without

454 spontaneously relaxing into a lower-energy state. For multiscale problems, the loop can

455 have a wide range of sizes, resulting in scale-free distributions alluded to earlier. In the

456 literature, the hysteretic nature of substorm is implicitly acknowledged (e.g., growth

457 phase vs expansion phase) but seldom emphasized. In our model, the energy storage and

458 release processes are governed by two clearly different processes (the storage represented

459 by the induction equation (2), and release process by current-driven instability and energy

460 redistribution, respectively). For studies of complex systems, explicit reference to

461 hysteresis is a needed step to conceptual clarity and quantitative treatment.

462    In terms of energy storage, the existing theories are biased toward producing large-

463 scale distributions rather than multi-scale ones. Consideration of a simple example

464 demonstrates the point. Suppose that the solar wind-magnetosphere interaction imposes a

465 boundary condition at the magnetopause. The distributions of pressure $p$ and magnetic

466 field $\mathbf{B}$ can be solved in principle. A general property of boundary-value problems of the

467     above sort is that small-scale features on the boundary decay quickly. Hence, one would

468     expect predominance of large-scale features in the CPS which is far away from the outer

469     magnetopause boundary. This expectation is inconsistent with the actual observation of

470     the CPS and the scale-free energy distribution which suggests a multiscale process at

471     play. In our model, energy is stored in multi-scale filamentary structures. As our

472     simulation showed, scale-free distributions resulted as a matter of course, without

473     appealing to extraneous factors or special circumstances.

474       The energy avalanche also warrants special attention. The traditional theory usually

475     invokes a substorm trigger at a special location, and the trigger excites a fast-mode MHD

476     wave that further disturbs the neighboring points (e.g., *Friedriech et al.*, 2000). While

477     similar to avalanche in appearance, the wave process implies that the expansion is at a

478     fixed speed, the pattern of propagation is regular (e.g., circular wave fronts), and the

479     reach of the expansion is global. In contrast, the avalanche model differs in these

480     important details. An avalanche occurs, in principle, in an irregular, often fractal area; the

481     network of nodes that are excited cannot be predicted beforehand, nor can the speed at

482     which the avalanche spreads on this network. Moreover, the avalanche can terminate at

483     any size; most in fact do not evolve into global events. This is the fundamental reason

484     why the avalanche model can naturally reproduce power-law distributions over energy,

485     size, and event time, while there is no such obvious path to scale-free distributions with

486     the traditional theory.

487       Finally, in the complexity paradigm, the exact nature or location of the trigger has

488     lesser import than in traditional models. Of course, the exact plasma physics that

489  contributes to the local instability which releases energy is important. What the above

490  statement alludes to, rather, is that the system's susceptibility to, global evolution, and

491  statistical properties of substorm may not be sensitive to the trigger. If a substorm is large,

492  it is likely due to the fact that the magnetic field structure out of which the substorm

493  erupts is more complex, rather than because it was triggered by a certain process. On a

494  more qualitative level, the present work argues for an important, if somewhat subtle

495  change of perspective. If a substorm is a global phenomenon, its underlying cause must

496  be global. The last snowflake that "triggers" a mountain avalanche is no different from

497  previous drops; it is thus incorrect to give it any special physical significance. The reason

498  why avalanches occur is that the overall snow cover has reached a critical state in a

499  global sense. This analogy encapsulates the point why trigger is not necessarily the

500  central problem in substorm. That the flu can trigger fatality is not a medically interesting

501  discovery; why the patient is susceptible to this trigger is. Similarly, the magnetotail has a

502  complex pattern of reaction to different disturbances (triggers). Most of these triggers do

503  not lead to a substorm. Those which do may not be fundamentally different from those

504  which do not. Therefore the study of substorm should be a study of how the magnetotail

505  behaves as a system, not merely about unstable modes which have a much higher

506  probability of occurrence, if not happening all the time.

507      Another new tapestry woven into the fabric of substorm theory is the role of the so-

508  called cross-scale coupling. The focus and forte of the traditional theory is transport

509  processes in the configurational (x) space. In this paper, our model was deliberately set

510  up so that it had no built-in structure in the initial current distribution, and a driver that

511    was also statistically constant and uncorrelated in space and time. Without any

512    preconditioning, the coupling of the two gave rise to a level of complexity that was not

513    anticipated. The physics behind these results is best elucidated in the Fourier-transformed

514    **k-**space.

515    Our results pointed to an interplay between flow **v** and current **j,** which may render the

516    debate about the primacy of one over the other a secondary issue, if not altogether

517    irrelevant. We demonstrated that a turbulent and spatially uncorrelated v can lead to

518    highly filamented current structures. In turn, a disruption in current j can set off

519    secondary flows, which helped unleash the avalanches.

520

521    **CONCLUSION**

522    Structuring of aurora is an unsolved problem important not only to magnetospheric

523    physics, but also to other problems of broad scientific interest. What we did in this paper

524    was not the provision of a solution, but a sketch that could help fashion a solution that

525    takes into account the fact that magnetospheric processes exhibits such complexity that

526    ideas and techniques developed in the study of nonlinear, non-equilibrium systems should

527    be used. Through simple but physically motivated argument and simulation, we have

528    explored an alternate view of energy storage and release in the CPS. This view

529    distinguishes itself from existing theoretical ideas in its emphasis of complexity and

530    reproduces several observed features which are mostly absent in traditional theories. The

531    highlights of our findings are:

532    1.   Turbulent magnetospheric convection creates elongated current filaments in the 533   central plasma sheet. The energy stored in these structures is multi-scale.

534   2.   The filaments have an arc-like appearance and may explain the formation of meso-

535        scale arcs reported by *Knudsen et al.* (2001);

536   3.   If the turbulence is strong enough or lasts long enough, the filamentary current

537        distribution reaches a criticality where energy avalanches are excited in the CPS;

538   4.   The distributions of avalanches over total released energy, peak power, and event

539        duration are scale-free. It is possible that phenomena we variously call substorms,

540        pseudo-breakups, saw-tooth events, etc, are subpopulations on this continuum

541        subjugate to common physics.

542   5.   There is a memory effect that governs the re-formation of filaments. An energy

543        avalanche does not completely erase the memory of current distribution preceding

544        the event. As a consequence, the remnant current distribution has a tendency to

545        replicate itself after the system starts the buildup phase again. This may explain

546        why auroral arcs tend to recur in the same general region of space.

547   These results hint strongly that energy storage and release processes in the magnetotail,

548   including the substorm, are multiscale involving both the classical cascade (which gives

549   rise to the turbulent flow) and inverse cascade featuring self-organization of small-scale

550   perturbations into larger-scale avalanches.

551

552   **Acknowledgments**. We thank Eric Donovan, David Knudsen, Jun Liang, Emma

553   Spanswick, Michel-Andre Vallieres-Nollet, and Tony Lui for discussions and helpful

554   comments pertaining to this study. This research was supported by the Canadian Space

555    Agency and Natural Sciences and Engineering Research Council of Canada.

556

653

654     Figure captions

655

656     Figure 1. Approximation of the magnetosphere (1a) as a collection of flux tubes moving

657     on a coupled lattice (1b). The motion is prescribed as a random, uncorrelated, and slow

658     shuffle to simulate the turbulent condition encountered in the central plasma sheet.

659

660     Figure 2. Time series of total magnetic energy stored on the lattice (top line) and energy

661     that is released through avalanche. Shown in the inset is a typical avalanche event and the

662     definition of total energy release ($E$), peak power ($P$), and event duration ($T$).

663

664     Figure 3. A typical avalanche event.

665

666     Figure 4. Probability density functions of energy release, peak power and event duration.

667     All three exhibit a power-law distribution suggesting scale-free dynamics.

668

669     Figure 5. Four examples of current distributions taken from the run in Figure 2. Plotted

670     alongside each distribution is the avalanche event it was in. The arrow in the plots on the

671     right-hand side indicates the exact moment when the current distribution was taken.

672

673     Figure 6. Current distributions from a different run of the model. The current distribution

674     is structurally different from Figure 5. Plot a is taken just before the onset of an avalanche,

675    and plot b right after. It can be seen that the avalanche does not completely remove the

676    memory the system has of the current distribution.

677

679

681

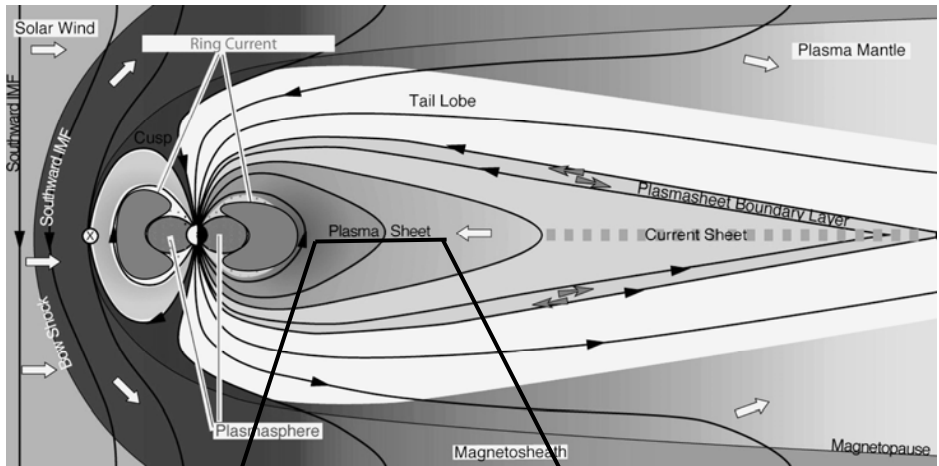

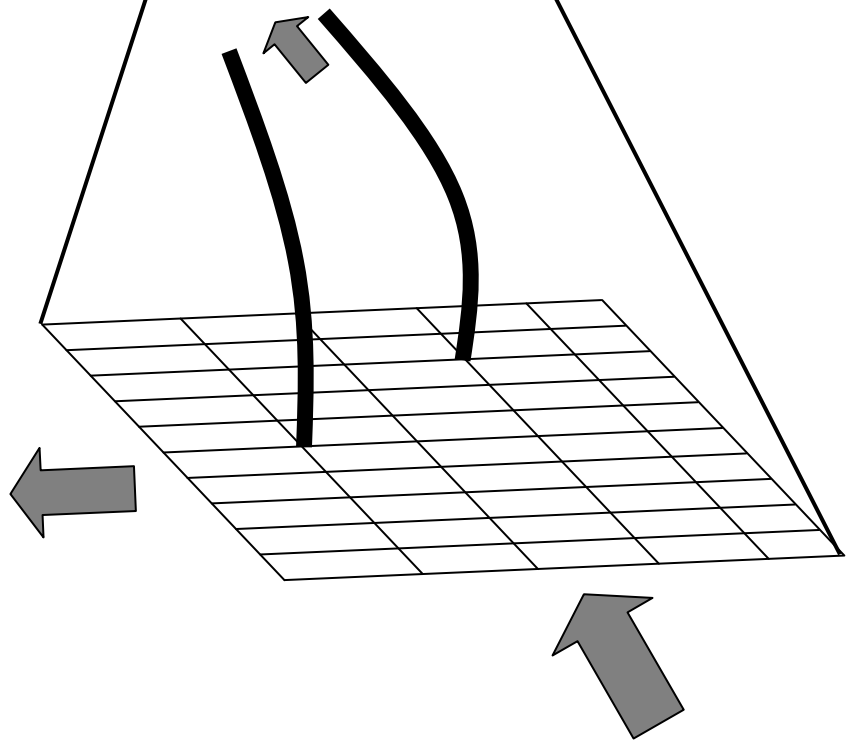

682

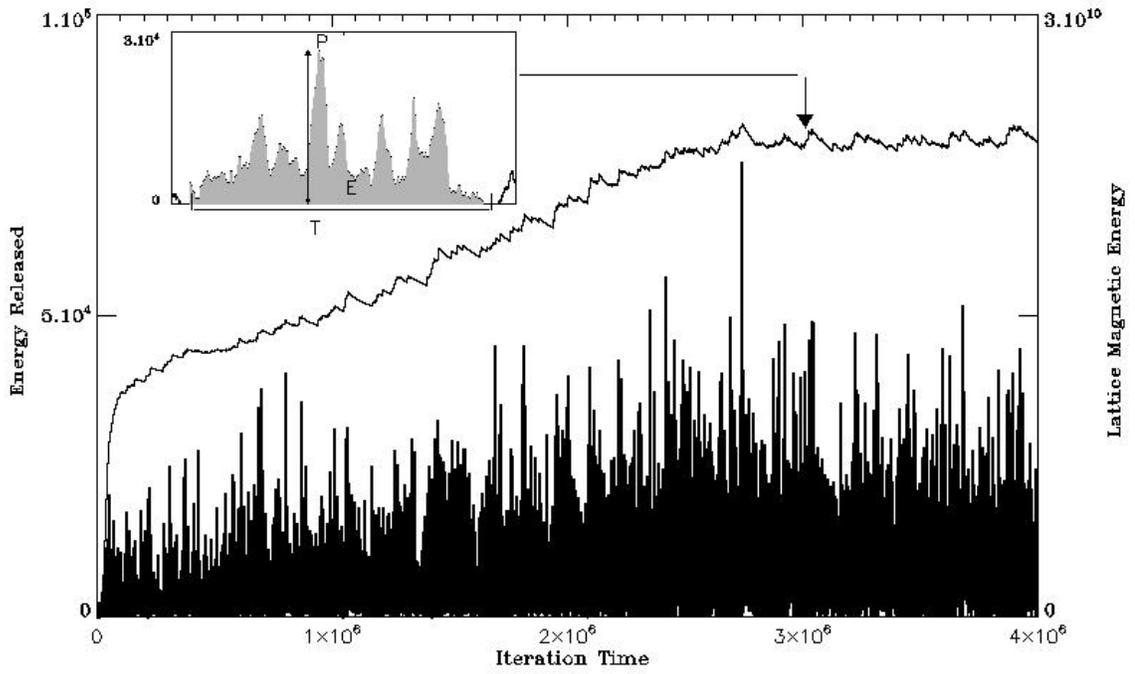

683
684

685

686

687

688

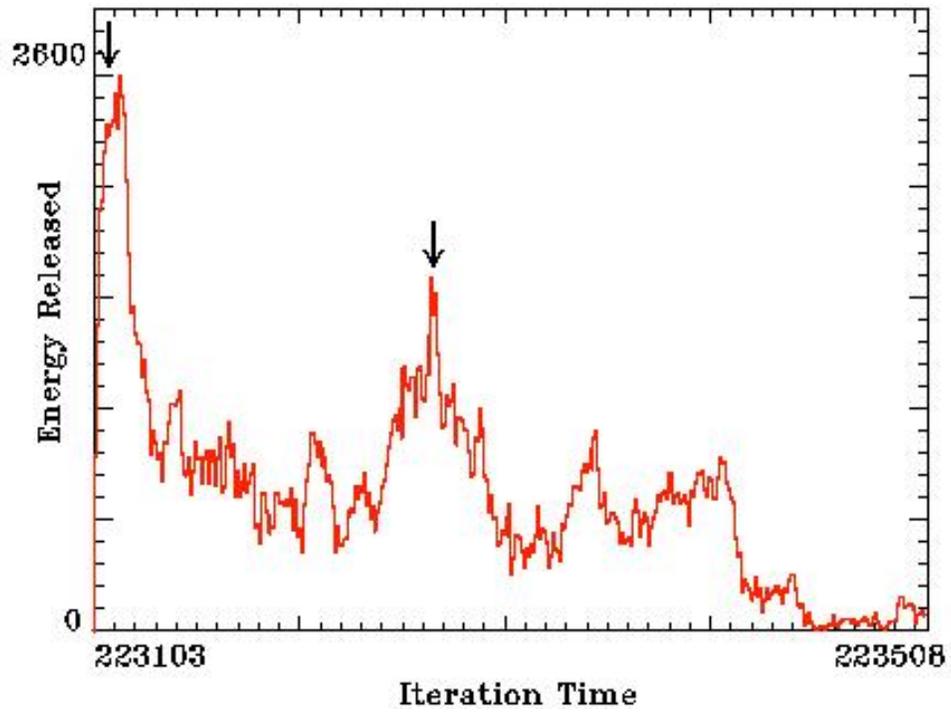

689

690

691

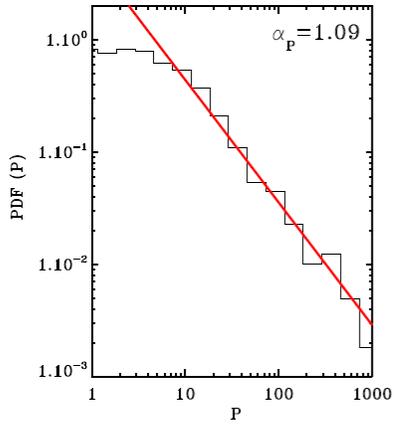 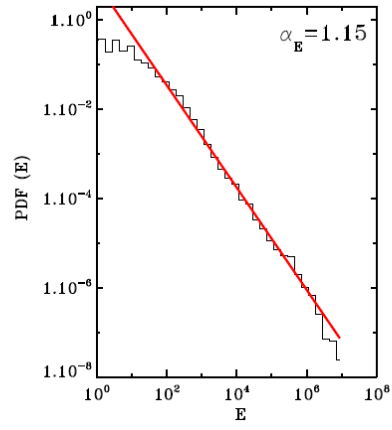 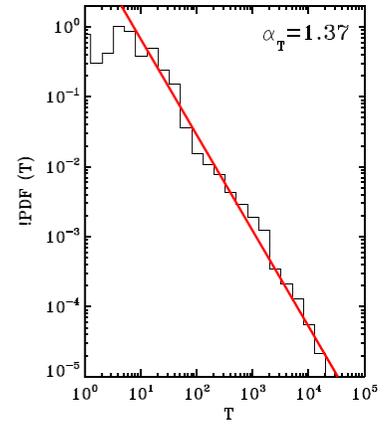

692

693

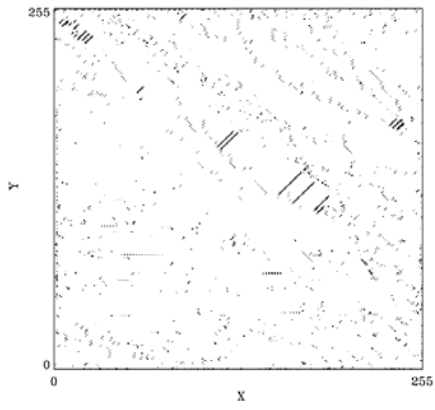
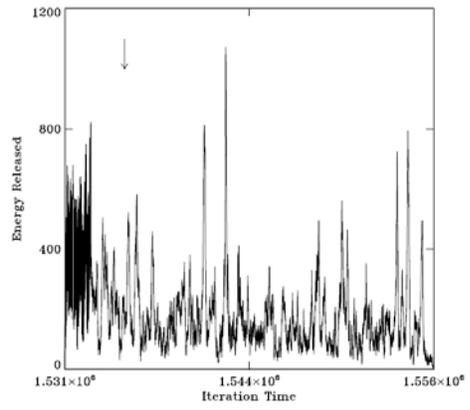

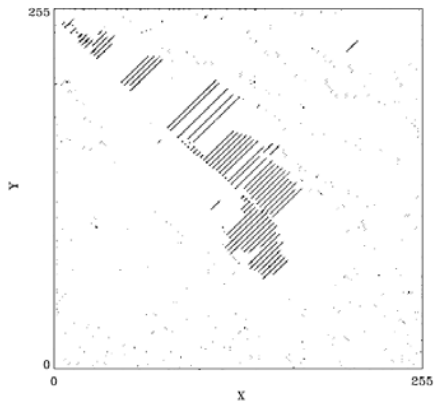
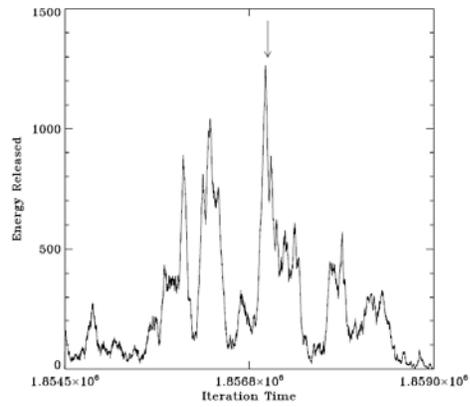

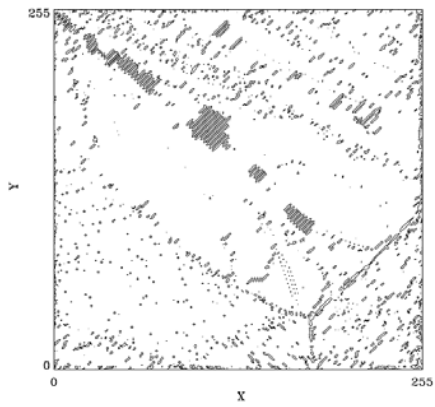
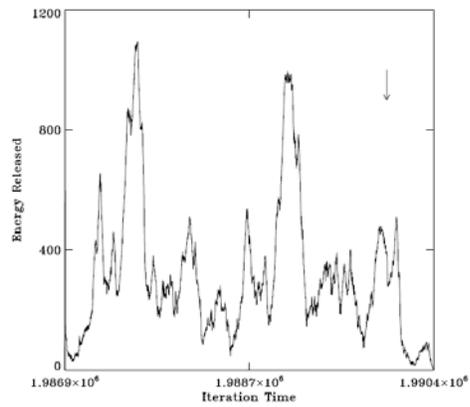

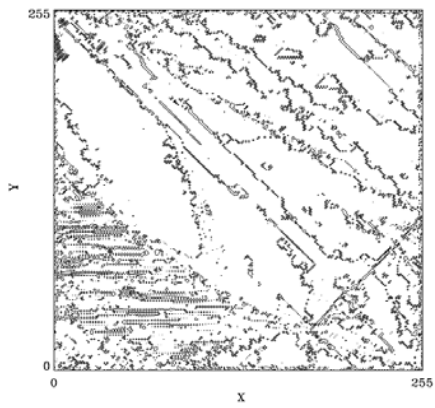
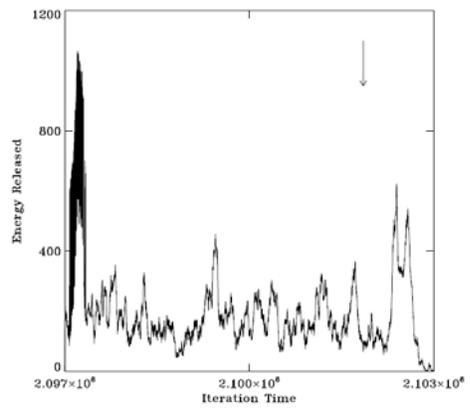

695

696

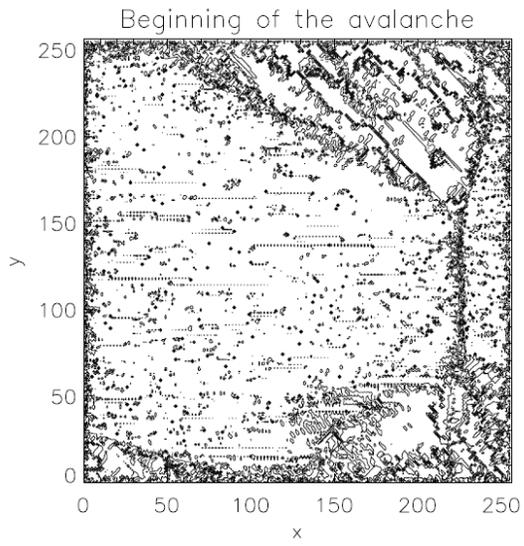 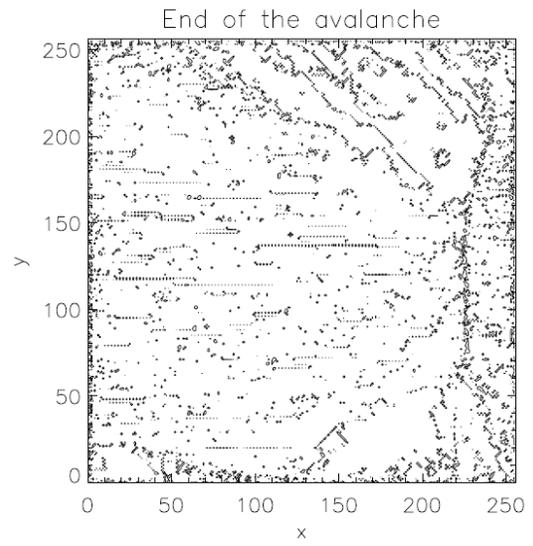

697

698